\documentclass[12pt]{article}

\begin{document}

\def\j{{\bf j}}
\def\tg{{ \rm tg}}


\vspace{5mm}

\begin{center}
{\Large \bf Possible quantum  kinematics. II. Non-minimal case}
\end{center}

\begin{center}
{\bf N.A. Gromov} \\
 Department of Mathematics, \\
Komi Science Center UrD RAS \\
Chernova st., 3a, Syktyvkar, 167982, Russia \\
E-mail: gromov@dm.komisc.ru
\end{center}

\begin{center}
{ \bf Abstract}
\end{center}

  The  quantum analogs of the $N$-dimensional 
Cayley-Klein spaces with different combinations of 
quantum and Cayley-Klein structures are described  for non-minimal multipliers, which include the first and the second powers  of contraction parameters in the transformation of deformation parameter.
The noncommutative  analogs of $(N-1)$-dimensional constant curvature spaces are introduced. 
Part of these spaces for $N=5$  are interpreted as the noncommutative analogs of $(1+3)$ space-time models.  As a result the  wide variety of the quantum deformations of realistic kinematics are suggested.


\section{Introduction}

Space-time is a fundamental conception that underlines the most significant physical theories. Therefore the analysis of a possible space-time models has the fundamental meaning for physics. 
In the previous papers \cite{GK-06},\cite{GK-04} the noncommutative  analogs of the possible commutative kinematics \cite{B-68}    
was constructed starting from the mathematical theory  of quantum groups and quantum vector spaces \cite{FRT}.
Cayley-Klein scheme of contractions and analytical continuations was applied to the five dimensional q-Euclidean space.
Different combinations of quantum structure
and Cayley-Klein scheme was described
with the help of permutations $\sigma. $

The transformation of the  deformation parameter $z=Jv$  under contraction is added in the quantum case as compared with the commutative one.
The analysis in  \cite{GK-06},\cite{GK-04} was confined to the minimal multiplier $J$, which has the first power multiplication of contraction parameters. 
This restriction imply 
that for certain combinations of quantum structure
and Cayley-Klein scheme kinematics contractions do
not exist. 
In order to all Cayley-Klein contractions
 was possible for all permutations $\sigma $
it is necessary to regard non-minimal multiplier $J$, which include the first and the second powers  of contraction parameters.
In this paper we find non-minimal multipliers for the general
quantum    Cayley-Klein  spaces (Sec. 2) and describe noncommutative four-dimensional space-time models (Sec. 3).

\section{Quantum    Cayley-Klein  spaces}

It was shown in the previous paper \cite{GK-06} that
the quantum N-dimensional  Cayley-Klein  space
$ O^N_v(j;\sigma;{\bf C}) $
is characterized by the following commutation relations of the Cartesian generators
      $$
\xi_{\sigma_{k}}\xi_{\sigma_{m}}  =
\xi_{\sigma_{m}} \xi_{\sigma_{k}} \cosh Jv -
i\xi_{\sigma_{m}} \xi_{\sigma_{k'}}
\frac{(1,{\sigma_{k'}})}{(1,{\sigma_{k}})}\sinh Jv, \;
k<m<{k'}, \;  k \not = m',
      $$
 \begin{equation}  
\xi_{\sigma_{k}}\xi_{\sigma_{m}}  =
\xi_{\sigma_{m}} \xi_{\sigma_{k}} \cosh Jv -
i\xi_{\sigma_{m'}} \xi_{\sigma_{k}}
\frac{(1,{\sigma_{m'}})}{(1,{\sigma_{m}})} \sinh Jv, \;
m'<k<m, \;  k \not = m',
\label{15-1}
 \end{equation}  
      $$
 [ \xi_{\sigma_{k}},\xi_{\sigma_{k'}}  ]
=  2i\epsilon
\sinh({{Jv} \over {2}}){(\cosh Jv)}^{n-k} \xi^2_{\sigma_{n+1}}
 \displaystyle{ {(1,{\sigma_{n+1}})^{2}}\over
{(1,{\sigma_{k}})(1,{\sigma_{k'}}) } } +
      $$
\begin{equation}
+ i\displaystyle{ {\sinh(Jv)} \over{(\cosh Jv)}^{k+1}
(1,{\sigma_{k}})(1,{\sigma_{k'}}) }
    \sum^n_{m=k+1}
   {(\cosh   Jv)}^m  ((1,{\sigma_{m}})^{2}\xi^2_{\sigma_{m}} +
(1,{\sigma_{m'}})^{2} \xi^2_{\sigma_{m'}}),
\label{15}
 \end{equation}
where $k,m=1,2,\ldots,n,\; N=2n+1$ or $N=2n,\; k'=N+1-k $,
permutation
$\sigma=(\sigma_1,\ldots,\sigma_N)$
describes definite combination of the quantum group structure and Cayley-Klein scheme of group contraction.
The invariant form under the coaction of the corresponding quantum orthogonal group on the Cayley-Klein space
$ O^N_v(j;\sigma;{\bf C}) $  is written as
$$
inv(j) =
\cosh (Jv{ \rho_1})\Biggl(  \epsilon (1,\sigma_{n+1})^2\xi^2_{\sigma_{n+1}}
{{(\cosh Jv)^{n}} \over {\cosh(Jv/2)}} +
$$
\begin{equation}
+ \sum^n_{k=1} ( (1,\sigma_{k})^{2}\xi^2_{\sigma_{k}} +
(1,\sigma_{k'})^{2}\xi^2_{\sigma_{k'}}) {(\cosh Jv)}^{k-1}\Biggr).
\label{16}
\end{equation}

The multiplier $J$ in the transformation  $z=Jv$ of the deformation parameter need be  chosen in such a way that all
indefinite relations in commutators, which appear under  nilpotent values of the contraction parameters are  canceled . As far as multipliers $(1,\sigma_k)$ and $(1,\sigma_{k'})$ enter symmetrically into the commutators (\ref{15-1}),(\ref{15}), we can put
$\sigma_k < \sigma_{k'} $ without  loss of  generality.
Then indefinite relations in commutators (\ref{15-1}) take the form $(1,\sigma_k)(1,\sigma_{k'})^{-1}=(\sigma_k,\sigma_{k'})$, where $k=1,2,\ldots,n $ for $N=2n+1$ and $k=1,2,\ldots,n-1 $ for $N=2n$. They are eliminated by the multiplier
\begin{equation}
 J_0=\displaystyle{\bigcup^n_{k=1}(\sigma_k,\sigma_{k'})}.
\label{16-5}
\end{equation}
It has the first power multiplication of contraction parameters and is the {\it minimal} multiplier, which guarantees
the existence of the Hopf algebra structure for the associated quantum group $SO_v(N;j;\sigma).$ The analysis in the previous paper \cite{GK-06} was confined to this minimal case.

The ``union'' of two multipliers
is understood as the 
multiplication of all parameters
$j_k,$ which occur  at least in one multiplier
  and the power of $j_k$ in the ``union'' is equal to its maximal power  in both multipliers,
for example, $(j_1j_2^2) \bigcup (j_2j_3)=j_1j_2^2j_3.$

If we take into account indefinite relations in commutators (\ref{15}), then we come to the non-minimal multiplier $J$, which consist of contraction parameters in the first and the second powers.
The indefinite relations in commutators (\ref{15}) have  the form
\begin{equation}
   \frac{\sum^n_{m=k+1} [(1,\sigma_{m})^{2} +
(1,\sigma_{m'})^{2}]}{(1,\sigma_k)(1,\sigma_{k'})}=
\frac{\sum^n_{m=k+1} (1,\sigma_{m})^{2}}{ (1,\sigma_k)^2(\sigma_k,\sigma_{k'})},
\label{16-1}
\end{equation}
$k=1,\ldots,n-1$ for even $ N=2n$ and
\begin{equation}
   \frac{(1,\sigma_{n+1}) + \sum^n_{m=k+1} [(1,\sigma_{m})^{2} +
(1,\sigma_{m'})^{2}]}{(1,\sigma_k)(1,\sigma_{k'})}=
\frac{(1,\sigma_{n+1}) + \sum^n_{m=k+1} (1,\sigma_{m})^{2}}{ (1,\sigma_k)^2(\sigma_k,\sigma_{k'})},
\label{16-2}
\end{equation}
$k=1,\ldots,n$ for odd $ N=2n+1$.
Let us introduce numbers
$$
i_k=\mbox{min}\{\sigma_{k+1},\ldots,\sigma_{n}\}, \;\; 
$$
then $k$th expression in (\ref{16-1}) or (\ref{16-2}) is equal to
$$ 
   \frac{ (1,i_{k})^{2}}{ (1,\sigma_k)^2(\sigma_k,\sigma_{k'})}=\left\{
     \begin{array}{cc}
    (i_k,\sigma_k)^{-2}(\sigma_k,\sigma_{k'})^{-1},  & i_k < \sigma_k  \\
      (\sigma_k,i_{k})(i_k,\sigma_{k'})^{-1}, & \sigma_k  < i_k < \sigma_{k'}\\
    (\sigma_k,\sigma_{k'})(\sigma_{k'},i_k)^2,  & i_k > \sigma_{k'}
   \end{array}
   \right.
$$  
 and compensative multiplier  for this expression is as follows
\begin{equation}
 J_1^{(k)}=\left\{
     \begin{array}{cc}
(i_k,\sigma_k)^{2}(\sigma_k,\sigma_{k'}), & i_k < \sigma_k  \\
     (i_k,\sigma_{k'}), & \sigma_k  < i_k < \sigma_{k'}\\
  1,  & i_k > \sigma_{k'}.
   \end{array}
   \right.
    \label{16-3}
\end{equation}
For all expressions in (\ref{16-1}) or (\ref{16-2}) compensative multiplier $J_1$ is obtained by the union         $ J_1=\bigcup_{k} J_1^{(k)}$.
Therefore the non-minimal  multiplier $J$ in the transformation  $z=Jv$ of the deformation parameter is given by
\begin{equation}
J=J_0\bigcup J_1=J_0 \bigcup_{k} J_1^{(k)}
\label{16-4}
\end{equation}
and include the first and the second powers of contraction parameters.

The quantum orthogonal Cayley-Klein sphere $S_v^{(N-1)}(j;\sigma) $ is obtained
as the quotient of $O_v^N(j;\sigma) $ by $ inv(j)=1. $
The quantum analogs of the intrinsic Beltrami coordinates on this quantum sphere
are given by  the sets of independent right or left generators
\begin{equation}
r_{\sigma_{i}-1}=\xi_{\sigma_i} \xi_1^{-1}, \quad
\hat{r}_{\sigma_{i}-1}=\xi_1^{-1} \xi_{\sigma_i} ,
\quad i=1,\ldots ,N, \quad i\neq k, \quad \sigma_k =1.
\label{17}
\end{equation}
Quantum kinematics are realized as quantum orthogonal sphere for $N=5$ and certain  values of contraction parameters.

\section{Quantum kinematics}
Quantum vector spaces $O^5_v(\j;\sigma)$ are generated by
$\xi_{\sigma_l},\; l=1,\ldots,5$ with commutation relations $(k=2,3,4)$
      $$
\xi_{\sigma_1}\xi_{\sigma_k} = \xi_{\sigma_k}\xi_{\sigma_1} \cosh(Jv)
-i\xi_{\sigma_k}\xi_{\sigma_5}{{(1,\sigma_5)} \over {(1,\sigma_1)}} \sinh(Jv), \;
     $$
     $$
\xi_{\sigma_k}\xi_{\sigma_5} = \xi_{\sigma_5}\xi_{\sigma_k} \cosh(Jv)
 - i\xi_{\sigma_1}\xi_{\sigma_k}{{(1,\sigma_1)} \over {(1,\sigma_5)}} \sinh(Jv),
\;\;
      $$
      $$
\xi_{\sigma_2}\xi_{\sigma_3} = \xi_{\sigma_3}\xi_{\sigma_2} \cosh(Jv)
-i\xi_{\sigma_3}\xi_{\sigma_4}{{(1,\sigma_4)} \over {(1,\sigma_2)}} \sinh(Jv), \;
      $$
      $$
\xi_{\sigma_3}\xi_{\sigma_4} = \xi_{\sigma_4}\xi_{\sigma_3} \cosh(Jv)
 - i\xi_{\sigma_2}\xi_{\sigma_3}{{(1,\sigma_2)} \over {(1,\sigma_4)}} \sinh(Jv),
      $$
      $$
\left  [\xi_{\sigma_2},\xi_{\sigma_4}  \right  ]  =
2i\xi_{\sigma_3}^2{{(1,\sigma_3)^2} \over {(1,\sigma_2)(1,\sigma_4)}} \sinh(Jv/2),
      $$
$$
 [ \xi_{\sigma_1},\xi_{\sigma_5}  ] =
2i\Biggl( \xi_{\sigma_3}^2 (1,\sigma_3)^2 \cosh(Jv) +
( \xi_{\sigma_2}^2 (1,\sigma_2)^2 +
$$
\begin{equation}
+\xi_{\sigma_4}^2 (1,\sigma_4)^2 )\cosh(Jv/2)  \Biggr)
{{\sinh(Jv/2)}\over {(1,\sigma_1)(1,\sigma_5)}}.
\label{1q}
\end{equation}

 By the  analysis of the multiplier (\ref{16-4}) for $N=5$
and commutation relations (\ref{1q}) of the quantum  space generators
we have find three 
different non-minimal $J$,
 namely $J=j_1^2j_2$, 
 $J=j_1j_2^2$
and  $J=j_1^2j_2^2$.
In the first two cases all different permutations give in result equivalent kinematics, so we can regard  only one permutation, say $ \hat{\sigma}=(2,1,3,4,5)$ for $J=j_1^2j_2$ and $ \check{\sigma} =(1,3,2,4,5) $ for $J=j_1j_2^2$. In the case of   $J=j_1^2j_2^2$ there are four nonequivalent quantum kinematics  corresponding to the following
permutations: 
$  
\tilde{\sigma} =(2,3,1,4,5),\;
\sigma_{I} =(3,1,5,2,4),\;
\sigma_{II} =(3,1,2,4,5),\;
\sigma_{III} =(3,2,1,4,5).
$

In order  to clarify the relation with the standard
In$\ddot{o}$n$\ddot{u}$--Wigner contraction procedure \cite{IW-53},
the mathematical parameter $j_1$
is replaced by  the physical one  $\tilde{j}_1T^{-1},$
and the  parameter  $j_2$ is replaced by   $ic^{-1},$ where
the standard  de Sitter kinematics $S_4^{(-)}$  with constant negative curvature is realized for $\tilde{j}_1=i$ and anti de Sitter kinematics $S_4^{(+)}$
with positive  curvature is obtained for $\tilde{j}_1=1.$

The limit $T \rightarrow \infty $ corresponds to the contraction  $j_1=\iota_1,$ and the limit $c \rightarrow \infty $ corresponds to
$j_2=\iota_2.$ The parameter $T$ is interpreted as
the curvature radius and has the  physical dimension of time
$[T]=[\mbox{time}],$ the parameter  $c$ is the light velocity
$[c]=[\mbox{length}][\mbox{time}]^{-1}. $
As far as  argument $Jv$ of hyperbolic function must be dimensionless, then  the physical dimension of deformation parameter is  equal to those of $J$, i.e.  $[v]=[J]$.
For $J=j_1^2j_2$ we obtain  $[v]=[cT^2]=[\mbox{length}][\mbox{time}]$,
for $J=j_1j_2^2$ we have $[v]=[c^2T]=[\mbox{length}]^2[\mbox{time}]^{-1}=[\mbox{length}][\mbox{velocity}]$
and for $J=j_1^2j_2^2$ the  physical dimension of the deformation parameter is given by
 $[v]=[c^2T^2]=[\mbox{length}]^2$.

The generator $\xi_1$ does not commute with others, therefore
it is  convenient to introduce right and left time
     $t=\xi_2\xi_1^{-1}, \hat{t}=\xi_1^{-1}\xi_2$ and space
     $r_k=\xi_{k+2}\xi_1^{-1}, \hat{r}_k=\xi_1^{-1}\xi_{k+2}, k=1,2,3,$
generators. The reason for this  definition is the simplification of expressions for commutation relations of
quantum kinematics. It is possible to use only, say, right generators,
but its commutators  are cumbersome in the case of the (anti) de Sitter kinematics.

\subsection{Quantum (anti) de Sitter kinematics}

For $\hat{\sigma}=(2,1,3,4,5)$ the mathematical multiplier $J=j_1^2j_2$ is rewritten in the form
$J=i\frac{\tilde{j_1}^2}{cT^2}=i\hat{J}$. Commutation relations of space and time generators of the (anti) de Sitter kinematics are
$$
S_v^{4(\pm)}(\hat{\sigma})=\Biggl\{t,{\bf r}|\;\;
\hat{t}r_p=\hat{r}_p\left(t \cos \hat{J}v +i
r_3 {\frac{1}{c}} \sin\hat{J}v\right),
$$
$$
\hat{t}r_3-\hat{r}_3t=2i \Biggl(
-\frac{\tilde{j}_1^2}{c^2T^2}\hat{r}_1r_1\cos\hat{J}v +
\left(1-\frac{\tilde{j}_1^2}{c^2T^2}\hat{r}_2r_2\right)
\cos \frac{\hat{J}v}{2}\Biggl)
\frac{cT^2}{\tilde{j}_1^2}\sin \frac{\hat{J}v}{2},
$$
\begin{equation} 
\hat{r}_pr_3=\left(\hat{r}_3 \cos\hat{J}v -i
\hat{t}  c \sin\hat{J}v \right)r_p, \;\;
\hat{r}_1r_2=\left(\hat{r}_2 \cos\hat{J}v -i\hat{t} \frac{cT}{\tilde{j_1}}
 \sin\hat{J}v \right)r_1 \Biggl\}.
\label{22-1}
\end{equation}
The right and left generators are connected as follows
$$
\hat{t}=t\cos\hat{J}v +i   r_3 {\frac{1}{c}}\sin\hat{J}v, \;\;
r_1=\hat{r}_1 \left(\cos\hat{J}v  + i r_2 \frac{\tilde{j}_1}{cT}\sin\hat{J}v \right) ,
$$
\begin{equation}
\hat{r}_2-r_2=2i\frac{\tilde{j_1}}{cT} \hat{r}_1r_1 \sin \frac{\hat{J}v}{2},\;\;
\hat{r}_3=r_3 \cos\hat{J}v  +itc \sin\hat{J}v .
\label{23-1}
\end{equation}

For $\check{\sigma}=(1,3,2,4,5)$ the mathematical multiplier $J=j_1j_2^2$ is rewritten in the form
$J=-\frac{\tilde{j_1}}{c^2T}=-\check{J}$. Commutation relations of space and time generators of the (anti) de Sitter kinematics are
$$
S_v^{4(\pm)}(\check{\sigma})=\Biggl\{t,{\bf r}|\;\;
\hat{r_1}t=\hat{t}\left(r_1 \cosh \check{J}v +
ir_2  \sinh \check{J}v\right),
$$
$$
\hat{t}r_2=\left(\hat{r}_2 \cosh \check{J}v +
i\hat{r}_1  \sinh \check{J}v\right)t,\;\;
\hat{t}r_3=\left(\hat{r}_3 \cosh \check{J}v + \frac{cT}{\tilde{j_1}}
  \sinh \check{J}v\right)t,
  $$
 \begin{equation}
\hat{r_1}r_2-r_2\hat{r}_1=2i \hat{t}tc^2\sinh \frac{\check{J}v}{2},\;\;
\hat{r}_pr_3=\left(\hat{r}_3 \cosh \check{J}v + \frac{cT}{\tilde{j_1}} \sinh \check{J}v \right)r_p \Biggl\}.
\label{22-2}
\end{equation}
The right and left generators are connected as follows
$$
t=\hat{t} \left( \cosh \check{J}v - \frac{\tilde{j_1}}{cT}  r_3 \sinh \check{J}v \right), \;\;
r_p=\hat{r}_p \left(\cosh \check{J}v  - \frac{\tilde{j_1}}{cT}  r_3 \sinh \check{J}v \right),
$$
\begin{equation}
\hat{r}_3-r_3=\frac{2}{\tilde{j_1}}cT \left( \hat{t}t \cosh \check{J}v  -\frac{\tilde{j_1}^2}{c^2T^2}\left(\hat{r}_1r_1+\hat{r}_2r_2 \right) \cosh \frac{\check{J}v}{2} \right)\sinh \frac{\check{J}v}{2}.
\label{23-2}
\end{equation}

For $\tilde{\sigma}=(2,3,1,4,5)$ the mathematical multiplier $J=j_1^2j_2^2$ is rewritten in the form
$J=-\frac{\tilde{j_1}^2}{c^2T^2}=-\tilde{J}$. Commutation relations of space and time generators of the (anti) de Sitter kinematics are
$$
S_v^{4(\pm)}(\tilde{\sigma})=\Biggl\{t,{\bf r}|\;\;
\hat{t}r_p=\hat{r}_p \left(t \cosh \tilde{J}v -
r_3 {\frac{1}{c}} \sinh \tilde{J}v \right),
$$
$$
\hat{t}r_3-\hat{r}_3t=-2 \Biggl[
\cosh \tilde{J}v - {\frac{\tilde{j}_1^2}{c^2T^2}}(\hat{r}_1r_1+\hat{r}_2r_2)
\cosh {\frac{\tilde{J}v}{2}} \Biggr] {\frac{cT^2}{\tilde{j}_1^2}}
\sinh {\frac{\tilde{J}v}{2}},
$$
\begin{equation}
\hat{r}_pr_3=\left( \hat{r}_3 \cosh \tilde{J}v
+\hat{t}  c \sinh \tilde{J}v \right)r_p, \;\;
\hat{r}_1r_2-\hat{r}_2r_1=-2i{\frac{c^2T^2}{\tilde{j}_1^2}} \sinh {\frac{\tilde{J}v}{2}}  \Biggl\},
\label{22}
\end{equation}
where  $k=1,2,3, \; p=1,2$. The left and right generators are connected by the   relations
$$
\hat{t}=t\cosh \tilde{J}v - r_3 {\frac{1}{c}} \sinh \tilde{J}v, \;\;
\hat{r}_1=r_1\cosh \tilde{J}v  + i r_2  \sinh \tilde{J}v,
$$
\begin{equation}
\hat{r}_2=r_2\cosh \tilde{J}v  - i r_1  \sinh \tilde{J}v,\;\;
\hat{r}_3=r_3\cosh \tilde{J}v - t c  \sinh \tilde{J}v.
\label{23}
\end{equation}

For $\sigma_{I}=(3,1,5,2,4)$  commutation relations of space and time generators of the (anti) de Sitter kinematics are
$$
S_v^{4(\pm)}(\sigma_{I})=\Biggl\{t,{\bf r}|\;\;
\hat{r}_1 t=\hat{t} \left(r_1 \cosh \tilde{J}v +
ir_2  \sinh \tilde{J}v \right),
$$
$$
\hat{t}r_2=\left(\hat{r}_2  \cosh \tilde{J}v +
i\hat{r}_1  \sinh \tilde{J}v \right),\;
\hat{r}_3 t=\left(\hat{t}  \cosh \tilde{J}v +
i\frac{T}{\tilde{j}_1} \sinh \tilde{J}v \right)r_3,
$$
$$
\hat{r}_1 r_3=\hat{r}_3 \left(r_1 \cosh \tilde{J}v +
ir_2  \sinh \tilde{J}v \right),\;
\hat{r}_3 r_2= \left(\hat{r}_2 \cosh \tilde{J}v +
i\hat{r}_1  \sinh \tilde{J}v \right)r_3,
$$
\begin{equation}
\hat{r}_1r_2-\hat{r}_2r_1=
2i \Biggl[-\frac{\tilde{j}_1^2}{c^2T^2}\hat{r}_3r_3
\cosh \tilde{J}v + \left( 1+\frac{\tilde{j}_1^2}{T^2}\hat{t}t \right) \cosh {\frac{\tilde{J}v}{2}} \Biggr]
\frac{c^2T^2}{\tilde{j}_1^2}
\sinh {\frac{\tilde{J}v}{2}} \Biggl\}.
\label{22-I}
\end{equation}
 The left and right generators are connected by the   relations
$$
\hat{r}_1=r_1\cosh \tilde{J}v  + i r_2  \sinh \tilde{J}v,\;\;
r_2=\hat{r}_2\cosh \tilde{J}v  + i \hat{r}_1  \sinh \tilde{J}v,
$$
\begin{equation}
r_3=\hat{r}_3 \left(\cosh \tilde{J}v +i \frac{\tilde{j}_1}{T}t \sinh \tilde{J}v \right),\;\;
t-\hat{t}=2i\hat{r}_3r_3\frac{\tilde{j}_1}{c^2T} 
\sinh \frac{\tilde{J}v}{2}.
\label{23-I}
\end{equation}

For $\sigma_{II}=(3,1,2,4,5)$  commutation relations of space and time generators of the (anti) de Sitter kinematics are
$$
S_v^{4(\pm)}(\sigma_{II})=\Biggl\{t,{\bf r}|\;\;
\hat{r}_1 t=\hat{t} \left(r_1 \cosh \tilde{J}v +
ir_3  \sinh \tilde{J}v \right),
$$
$$
\hat{t}r_2=\left(\hat{r}_2  \cosh \tilde{J}v +
i\frac{cT}{\tilde{j}_1} \sinh \tilde{J}v \right)t,\;
\hat{t}r_3=\left(\hat{r}_3  \cosh \tilde{J}v +
i\hat{r}_1 \sinh \tilde{J}v \right)t,
$$
$$
\hat{r}_1 r_2=\hat{r}_2 \left(r_1 \cosh \tilde{J}v +
ir_3  \sinh \tilde{J}v \right),\;
\hat{r}_2 r_3= \left(\hat{r}_3 \cosh \tilde{J}v +
i\hat{r}_1  \sinh \tilde{J}v \right)r_2,
$$
\begin{equation}
\hat{r}_1r_3-\hat{r}_3r_1=
2i \Biggl[\frac{\tilde{j}_1^2}{T^2}\hat{t}t\cosh \tilde{J}v + \left( 1-\frac{\tilde{j}_1^2}{c^2T^2}\hat{r}_2r_2 \right) 
\cosh {\frac{\tilde{J}v}{2}} \Biggr]
\frac{c^2T^2}{\tilde{j}_1^2}
\sinh {\frac{\tilde{J}v}{2}} \Biggl\}.
\label{22-II}
\end{equation}
 The left and right generators are connected by the   relations
$$
\hat{r}_1=r_1\cosh \tilde{J}v  + i r_3  \sinh \tilde{J}v,\;\;
\hat{r}_2=r_2\cosh \tilde{J}v  + 2\hat{t}t\tilde{j}_1\frac{c}{T}  \sinh \frac{\tilde{J}v}{2},
$$
\begin{equation}
r_3=\hat{r}_3 \cosh \tilde{J}v +i \hat{r}_1 \sinh \tilde{J}v,\;\;
t=\hat{t} \cosh \tilde{J}v - \hat{t}r_2\frac{\tilde{j}_1}{cT}
 \sinh \tilde{J}v.
\label{23-II}
\end{equation}

For $\sigma_{III}=(3,2,1,4,5)$  commutation relations of space and time generators of the (anti) de Sitter kinematics are
$$
S_v^{4(\pm)}(\sigma_{III})=\Biggl\{t,{\bf r}|\;\;
\hat{r}_1 t=\hat{t} \left(r_1 \cosh \tilde{J}v +
ir_3  \sinh \tilde{J}v \right),
$$
$$
\hat{t}r_2 -\hat{r}_2t= -2\frac{cT^2}{\tilde{j}_1^2}
 \sinh \tilde{J}v ,\;
\hat{t}r_3=\left(\hat{r}_3  \cosh \tilde{J}v +
i\hat{r}_1 \sinh \tilde{J}v \right)t,
$$
$$
\hat{r}_1 r_2=\hat{r}_2 \left(r_1 \cosh \tilde{J}v +
ir_3  \sinh \tilde{J}v \right),\;
\hat{r}_2 r_3= \left(\hat{r}_3 \cosh \tilde{J}v +
i\hat{r}_1  \sinh \tilde{J}v \right)r_2,
$$
\begin{equation}
\hat{r}_1r_3-\hat{r}_3r_1=
2i \Biggl[\cosh \tilde{J}v + \frac{\tilde{j}_1^2}{T^2}\left( \hat{t}t -\frac{1}{c^2}\hat{r}_2r_2 \right)
\cosh {\frac{\tilde{J}v}{2}} \Biggr]
\frac{c^2T^2}{\tilde{j}_1^2}
\sinh {\frac{\tilde{J}v}{2}} \Biggl\}.
\label{22-III}
\end{equation}
 The left and right generators are connected by the   relations
$$
\hat{r}_1=r_1\cosh \tilde{J}v  + i r_3  \sinh \tilde{J}v,\;\;
r_2=\hat{r}_2\cosh \tilde{J}v  + c\hat{t}\sinh \tilde{J}v, 
$$
\begin{equation}
r_3=\hat{r}_3 \cosh \tilde{J}v +i \hat{r}_1 \sinh \tilde{J}v,\;\;
\hat{t}=t \cosh \tilde{J}v - r_2\frac{1}{c}
 \sinh \tilde{J}v.
\label{23-III}
\end{equation}

\subsection{Quantum  Minkowski  kinematics}

 Zero curvature limit $ T \rightarrow \infty $ transform
 (anti) de Sitter kinematics to Minkowski one. For permutations $ \hat{\sigma}, \tilde{\sigma}, \sigma_{I}, \sigma_{II}, \sigma_{III} $ left and  right space-time generators coincide $\hat{t}=t, \hat{r}_k=r_k$ and commutation relations take the form
$$ 
M_v^4(\tilde{\sigma})=\Biggl\{t,{\bf r}|\;\;  [t,r_p]=0, \;[t,r_3]=-\frac{v}{c}, \;
[r_1,r_2]=-iv,\; [r_p,r_3]=0,\; p=1,2 \Biggl\},
$$ 
$$ 
M_v^4(\hat{\sigma})=\Biggl\{t,{\bf r}|\;\; [t,r_p]=0,\; [t,r_3]=iv, \;  \;
[r_i,r_k]=0, \;i,k=1,2,3  \Biggl\},
$$ 
$$ 
M_v^4(\sigma_{I})=\Biggl\{t,{\bf r}|\;\; [t,r_k]=0,\;
[r_1,r_3]=[r_2,r_3]=0, \; [r_1,r_2]=iv  \Biggl\},
$$ 
$$ 
M_v^4(\sigma_{II})=\Biggl\{t,{\bf r}|\;\; [t,r_k]=0,\;
[r_1,r_2]=[r_2,r_3]=0, \; [r_1,r_3]=iv  \Biggl\},
$$ 
$$ 
M_v^4(\sigma_{III})=\Biggl\{t,{\bf r}|\;\; [t,r_1]=[t,r_3]=0,\; [t,r_2]=-\frac{v}{c}, 
$$
\begin{equation}
[r_1,r_2]=[r_2,r_3]=0, \; [r_1,r_3]=iv  \Biggl\},
\label{24-III}
\end{equation}
For permutation $\check{\sigma}$ we have $\hat{t}=t, \hat{r}_p=r_p $, but $\hat{r}_3=r_3-\frac{v}{c}t^2$ and Minkowski kinematics is given by
$$
M_v^4(\check{\sigma})=\Biggl\{t,{\bf r}|\;\; [t,r_p]=0,\;
[t,r_3]=\frac{v}{c}t(1-t^2), \;
[r_1,r_2]=0, \;
$$
\begin{equation}
 [r_2,r_3]=\frac{v}{c}r_2(1-t^2), \;
[r_1,r_3]=\frac{v}{c}r_1(1-t^2)  \Biggl\}.
\label{24-2}
\end{equation}
$ M_v^4(\sigma_{I}) $ and $ M_v^4(\sigma_{II}) $ are connected by replacement  of space generators $r_2$ with $r_3$ and vice-versa, therefore they can be regarded as equivalent kinematics.
The same is true for $M_v^4(\tilde{\sigma})$ and $M_v^4(\sigma_{III})$.
so we obtain four nonequivalent Minkowski  kinematics.

\subsection{Quantum  Newton kinematics}

Newton kinematics are obtained from (anti) de Sitter kinematics by the non-relativistic limit $ c \rightarrow \infty $ 
 and  are characterized by the following commutation relations of generators
$$
N_v^{4(\pm)}(\tilde{\sigma})=\Biggl\{t,{\bf r} | \; [t,r_k]=0, \;\;
[r_1,r_2]=-iv,\;\;
[r_p,r_3]=0  \Biggl\},
$$
$$
N_v^{4(\pm)}(\check{\sigma})=\Biggl\{t,{\bf r} | \;
[t,r_k]=0, \;\;
[r_1,r_2]=i\tilde{j}_1\frac{v}{T}t^2, \;\;
[r_p,r_3]=0 \Biggl\},
$$
$$
N_v^{4(\pm)}(\sigma_{I})=\Biggl\{t,{\bf r} | \; [t,r_k]=0, \;
[r_1,r_3]=[r_2,r_3]=0,\;
[r_1,r_2]=iv\left(1+\frac{\tilde{j}_1^2}{T^2}t^2  \right)
  \Biggl\},
$$
$$
N_v^{4(\pm)}(\sigma_{II})=N_v^{4(\pm)}(\sigma_{III})=\Biggl\{t,{\bf r} | \; [t,r_k]=0, \; [r_1,r_2]=[r_2,r_3]=0,
$$
\begin{equation}
[r_1,r_3]=iv\left(1+\frac{\tilde{j}_1^2}{T^2}t^2  \right)
 \Biggl\}.
\label{25-0}
\end{equation}
Unlike previous cases, where $\hat{t}=t, \hat{r}_k=r_k$, for permutation $\hat{\sigma}$ the third left and right space generators are connected as
$\hat{r}_3=r_3 +i\tilde{j}_1^2\frac{v}{T^2}$, therefore commutation relations in this case are
$$
N_v^{4(\pm)}(\hat{\sigma})=\Biggl\{t,{\bf r} | \;
[t,r_p]=0, \; [t,r_3]=iv\left(1+\frac{\tilde{j}_1^2}{T^2}t^2\right),
$$
\begin{equation}
[r_1,r_2]=-i\tilde{j}_1\frac{v}{T}tr_1, \;
[r_p,r_3]=0 \Biggl\}.
\label{25-2}
\end{equation}
Nonequivalent (anti) de Sitter kinematics 
$ S_v^{4(\pm)}(\sigma_{II})$ and $ S_v^{4(\pm)}(\sigma_{III})$ become identical one 
$N_v^{4(\pm)}(\sigma_{II})=N_v^{4(\pm)}(\sigma_{III}) $
in the non-relativistic limit.
$ N_v^4(\sigma_{I}) $ and $ N_v^4(\sigma_{II}) $ are connected by replacement  of space generators $r_2$ with $r_3$ and vice-versa, therefore they can be regarded as equivalent kinematics. So there are four nonequivalent Newton kinematics.
\subsection{Quantum  Galilei  kinematics}

Galilei kinematics can be obtained by the non-relativistic limit $ c \rightarrow \infty $ of the Minkowski kinematics (\ref{24-III}),(\ref{24-2}) or by the zero curvature limit
$ T \rightarrow \infty $ of the Newton kinematics (\ref{25-0}),(\ref{25-2}). They have identical left and right space-time generators, which are commutative for
$G_v^{4}(\check{\sigma})$.
Galilei kinematics $G_v^{4}(\hat{\sigma})$  has one
nonzero commutator of time and space generators $[t,r_3]=iv$ as well as Minkowski kinematics $M_v^{4}(\hat{\sigma})$ Minkowski kinematics $M_v^{4}(\hat{\sigma})$(\ref{24-III}).
For Galilei kinematics 
$
G_v^{4}(\tilde{\sigma}), G_v^{4}(\sigma_{I}),
G_v^{4}(\sigma_{II}), G_v^{4}(\sigma_{III})
$ 
 there is only one nonzero commutator of space generators $[r_1,r_2]=\pm iv$ or $[r_1,r_3]=iv$, i.e. all these kinematics are physically equivalent.
So there  exist two nonequivalent noncommutative Galilei kinematics and for permutation $\check{\sigma}$ the non-relativistic and zero curvature limits give in result Galilei kinematics with totally commuting space-time generators.

\subsection{Quantum  Carroll kinematics}

The exotic Carroll kinematics   are also realized as constant curvature spaces for $j_2=j_3=1,\;j_4=\iota_4$, but with different interpretation of the
Beltrami coordinates, namely $r_k=\xi_{k+1} \xi_1^{-1},\;k=1,2,3$
are  the space generators and $t=\xi_5 \xi_1^{-1}$ is the time generator.
 Due to this  interpretation the new physical  dimensions of the contraction parameters  appear: the parameter $j_1$ is replaced by
$\tilde{j}_1R^{-1},$ where $R \rightarrow \infty $ corresponds to
$j_1=\iota_1$ and $[R]=[\mbox{length}] $.

The nonequivalent Carroll kinematics with non-minimal multiplier $J=j_1^2\iota_4$  are  achieved  for the three substitutions: $\hat{\sigma}=(2,1,3,4,5),$ $ \sigma''=(2,1,3,5,4),$ $\sigma'''=(2,3,1,5,4)$.
The  physical dimension of the deformation parameter is equal to  $[v]=[R^2]=[\mbox{length}]^2$.
The corresponding  kinematics are given by space-time generators 
with the commutation relations
\begin{equation}
C_v^{4(\pm)}(\hat{\sigma})=\Biggl\{t,{\bf r}|\;
[r_1,t]=iv \left (1+ {\frac{\tilde{j_1^2}}{R^2}} {\bf r}^2\right ),\; [t,r_2]=[t,r_3]=0,\;  [r_i,r_k]=0
 \Biggl\},
\label{26-1}
\end{equation}
where ${\bf r}^2=r_1^2+r_2^2+r_3^2$ with $\hat{r}_k=r_k$, but left and right time generators are connected as $\hat{t}=t+i\frac{\tilde{j}_1^2}{R^2}vr_1 $;
$$
C_v^{4(\pm)}(\sigma'')=
$$
\begin{equation}
=\Biggl\{t,{\bf r}|\;
[t,r_p]=iv  \frac{\tilde{j}_1^3}{R^3}r_2^2r_p, \; [t,r_2]=iv\frac{\tilde{j}_1}{R}\left(1+ \frac{\tilde{j}_1^2}{R^2}r_2^2 \right)r_2,\;  [r_i,r_k]=0
 \Biggl\},
\label{26-2}
\end{equation}
where $p=1,3,\;\hat{r}_k=r_k,\; \hat{t}=t-i\frac{\tilde{j}_1^3}{R^3}vr_2^2$ 
and
$$
C_v^{4(\pm)}(\sigma''')=\Biggl\{t,{\bf r}|\;
[t,r_k]=0,\;
[r_1,r_2]=iv \left[1+ \frac{\tilde{j}_1^2}{R^2}\left( t^2+r_1^2+r_3^2\right)\right], \; 
$$
\begin{equation}
[r_1,r_3]=[r_2,r_3]=0
 \Biggl\},
\label{26-3}
\end{equation}
with $\hat{r}_2=r_2+i\frac{\tilde{j}_1^2}{R^2} vr_1,\; \hat{r}_1=r_1, \hat{r}_3=r_3, \hat{t}=t$. 
The kinematics $C_v^{4(\pm)}(\hat{\sigma})$ and $C_v^{4(\pm)}(\sigma''')$ are transformed to each other by the replacement $ t \rightarrow r_2,\; r_2 \rightarrow t $, i.e. they are mathematically isomorphic, but physically nonequivalent.
So we obtain tree nonequivalent Carroll kinematics.

 Quantum analogs of the  zero curvature Carroll kinematics are obtained in the limit $R \rightarrow  \infty $. 
 For $C_v^{4(0)}(\sigma'')$ all generators are commute.
 For $C_v^{4(0)}(\hat{\sigma})$ the only nonzero commutator
 is $[r_1,t]=iv$ and for $C_v^{4(0)}(\sigma''')$ such commutator is $[r_1,r_2]=iv$.

 \section{Conclusion}

The quantum Cayley-Klein spaces of constant curvature $O_q^N(j;\sigma)$ are uniformly obtained from 
the quantum Euclidean space $O_q^N$ in Cartesian coordinates 
by the standard trick with real, complex, and nilpotent  numbers, using a q-analog of Beltrami coordinates.
The transformation of the quantum deformation parameter $Z=Jv$ under contraction is the important ingredient of the noncommutative quantum groups and noncommutative quantum spaces. Unlike previous papers on this subject 
\cite{GK-06},\cite{GK-04} second power of contraction parameters are included in the multiplier $J$, what make
all kinematics contractions 
admissible.
The different combinations of quantum structure
and Cayley-Klein scheme of contractions and analytical continuations are described
with the help of permutations $\sigma. $
As a result  the   quantum deformations of four dimensional (anti) de Sitter, Minkowski, Newton,  Galilei and Carroll kinematics with non-minimal transformations of deformation parameter are obtained.

We have found six types of the noncommutative realistic (anti) de Sitter kinematics with non-minimal multipliers, which admit all non-relativistic and zero curvature limits and three types of the nonequivalent Carroll kinematics.
For Minkowski and Newton kinematics there are four  types and
two types for noncommutative Galilei kinematics.
In addition every kinematics has two types with minimal multipliers \cite{GK-06}.
This demonstrate a wide variety of the quantum deformations of the commutative space-time models.
The remarkable property of the limit kinematics 
is that for some of them commutation relations of  generators are proportional to a {\it numbers} instead of generators, for example, all Minkowski 
(\ref{24-III}), Newton 
$N_v^{4(\pm)}(\tilde{\sigma})$ (\ref{25-0}), 
all Galilei 
and all flat Carroll kinematics. Moreover Galilei $G_v^{4}(\check{\sigma})$ and Carroll $C_v^{4(0)}(\sigma'')$ kinematics have commutative space-time generators.
In other words commutative Galilei kinematics or  
{\it simplest} deformations of  Galilei and Minkowski kinematics, when zero commutation relations of generators become proportional to a {\it numbers}, can be obtain from quantum (anti) de Sitter kinematics by contractions with non-minimal transformations of deformation parameter.

The quantum Galilei kinematics $G_v^4(\tilde{\sigma}), G_v^4(\sigma_{I}), G_v^4(\sigma_{II}), G_v^4(\sigma_{III}),$
 Newton $N_v^{4(\pm)}(\tilde{\sigma})$, Minkowski $M_v^4(\sigma_{I})$ and Carroll $C_v^{4(0)}(\sigma''')$ have the same nonzero commutator $[r_1,r_2]=iv$.
Equal  nonzero commutator $[t,r_3]=iv$ have Minkowski $M_v^4(\hat{\sigma})$,  Galilei $G_v^4(\hat{\sigma})$ and
Carroll $C_v^{4(0)}(\hat{\sigma})$ kinematics.
In spite of the fact that   commutation relations of generators of some kinematics are identical, these kinematics are physically different. Mathematically isomorphic kinematics may be physically nonequivalent.

\section{Acknowledgments}
This work was supported by the program "Fundamental problems of nonlinear dynamics" of the Russian Academy of Sciences.

\end{document}